# An Empirical-cum-Statistical Approach to Power-Performance Characterization of Concurrent GPU Kernels

Nilanjan Goswami, *Member, IEEE,* Yuhai Li, Amer Qouneh, Chao Li, *Member, IEEE*, Tao Li, *Senior Member, IEEE*

**Abstract**— Growing deployment of power and energy efficient throughput accelerators (GPU) in data centers demands enhancement of power-performance co-optimization capabilities of GPUs. Realization of exascale computing using accelerators requires further improvements in power efficiency. With hardwired kernel concurrency enablement in accelerators, inter- and intra-workload simultaneous kernels computation predicts increased throughput at lower energy budget. To improve Performance-per-Watt metric of the architectures, a systematic empirical study of real-world throughput workloads (with concurrent kernel execution) is required. To this end, we propose a multi-kernel throughput workload generation framework that will facilitate aggressive energy and performance management of exascale data centers and will stimulate synergistic power-performance co-optimization of throughput architectures. Also, we demonstrate a multi-kernel throughput benchmark suite based on the framework that encapsulates symmetric, asymmetric and co-existing (often appears together) kernel based workloads. On average, our analysis reveals that spatial and temporal concurrency within kernel execution in throughput architectures saves energy consumption by 32%, 26% and 33% in GTX470, Tesla M2050 and Tesla K20 across 12 benchmarks. Concurrency and enhanced utilization are often correlated but do not imply significant deviation in power dissipation. Diversity analysis of proposed multi-kernels confirms characteristic variation and power-profile diversity within the suite. Besides, we explain several findings regarding power-performance co-optimization of concurrent throughput workloads.

---◆---

## 1 INTRODUCTION

DUE to improved energy efficiency [1] and better performance [2-4], throughput computing architectures such as GPUs (Nvidia [5], AMD [6]), dedicated accelerators (Intel MIC [7]), IBM Cell processors [8] are increasingly adopted to solve high performance computing problems [9-14] in data centers [15-19] and supercomputers (Tianhe-1A, Nebulae, Tsubame). Besides, the price-to-performance ratio of throughput architecture has also been decreasing over the years. As a result, we see an increasing trend of GPU based *throughput-computing-in-the-cloud* infrastructure [19] that employs shared virtualized GPUs (better resource sharing, improved power efficiency) for throughput computing or non-virtualized dedicated GPUs attached to virtual machine instances to execute performance-critical HPC tasks using GPU *passthrough* [20].

To compensate single thread processing energy overhead, a large percentage of the parallel thread processing in throughput processors often shares hardware structures (shared memory, scheduler, issue/decode unit, etc.). As a result, overall energy efficiency enhances. Those processors try to achieve energy efficiency, concurrent execution paradigm, and performance improvement simultaneously. Myriad concurrency at various levels of throughput execution offers an order of magnitude performance enhancement. However, increased concurrency and performance do not always map to improved energy and power efficiency. Nevertheless, it is still possible to exploit further concurrency using new computational paradigms and architectural enhancements (such as concurrent kernels). Evidently, kernel level concurrency has significant implications on performance and power. However, a thorough exploration of power-performance characteristics of concurrent throughput kernels is still lacking. There is a need to identify the representative mix of workloads, which reduces overall energy footprint and retain throughput. To this end, we propose a flexible framework to mix emerging throughput workloads that can execute together to achieve improved performance-per-watt. Furthermore, we propose a workload consolidation framework for throughput architectures that is energy and performance optimized. Such exploration and framework might benefit HPC cloud architects and data center designers in many ways. The research helps in the selection of appropriate type/count of throughput processors (low-power, high-performance,

---

- *Nilanjan Goswami and Tao Li are with Department of Electrical and Computer Engineering at the University of Florida, USA, E-mail: taoli@ece.ufl.edu, nil@ieee.org.*
- *Chao Li is with the Department of Computer Science and Engineering at the Shanghai Jiao Tong University, China, E-mail: lichao@cs.sjtu.edu.cn.*
- *Amer Qouneh is with the Department of Computer Engineering at the Western New England Universit, USA, E-mail: aqouneh@ufl.edu.*
- *Yuhai Li is with the Xi'an Jiaotong University, China, E-mail: li-yuhai.cn@gmail.com.*

and energy efficient), power-performance co-optimization of throughput applications, restructuring cooling infrastructure to achieve improved perf/watt, optimizing power delivery network and data center network and much more.

HPC workloads often do not exhaust available resources (registers, shared memory, threads, thread blocks, constant/texture memory) in GPUs due to algorithmic limitation, performance target, energy capping, etc. Since performance is the most dominant driving force, previous works [21-25] have only addressed the concurrency performance correlation phenomenon. From compile time kernel fusion to runtime dynamic elastic kernels, all have discussed performance shortfall due to insufficient concurrency at various levels. To design an energy efficient architecture, throughput architects need to consider energy and power implications of kernel level concurrency as well. To answer such simple questions, we delve into the systematic exploration of throughput workloads that unleashes power-performance co-characterization. The main contributions of the paper are:

(a) For the first time, we propose and present a power-performance characterization of emerging multi-kernel throughput workload suite.

(b) We also propose a systematic methodology for multi-kernel throughput workload creation for emerging systems.

(c) Furthermore, we analyze and implement a multi-kernel benchmark generation framework.

(d) We have thoroughly evaluated and analyzed the synergistic optimization of multi-kernel workloads regarding power, performance, and utilization. Furthermore, we present several findings from the multi-kernel workload characterization.

To the best of our knowledge, this is first work on power analysis of multi-kernel throughput workloads. The rest of the paper organize as follows: Section 2 provides motivations behind the work (2.1) and provides necessary background (2.2), Section 3 proposes the methodology, Section 4 explains the experimental setup, Section 5 analyses the benchmark suite, Section 6 highlights related research and finally Section 7 concludes the paper.

## 2 MOTIVATION AND BACKGROUND

### 2.1 Improved Power, Performance, and Utilization Efficiency

Emerging throughput architectures and workloads are continuously evolving. Cloud-based HPC datacenters are gradually procuring GPU-like throughput processors for energy-efficient compute-intensive tasks. In addition to energy efficiency, improved utilization and performance are also critical in amortizing the long-term data center operating cost. To improve the efficiency of the emerging throughput architectures, they offer various levels of concurrency (application, kernel, thread, and data). However, those workloads and architectures still lack a thorough and in-depth investigation of power-performance co-optimization opportunities at the application, job, and task levels. For example, what is the most appropriate combination of throughput kernels that can provide energy efficient execution and minimize performance degradation? Is there any optimum systematic approach for selecting and overlapping throughput workloads? How do workload characteristics and power profile impact co-optimization of overlapped kernel execution? What amount of underutilized resources can we still leverage in improving energy efficiency? During simultaneous execution of multiple kernels, is it possible to gain collective power-performance improvement by sacrificing power/performance of individual kernels? In this work, we seek answers to these questions.

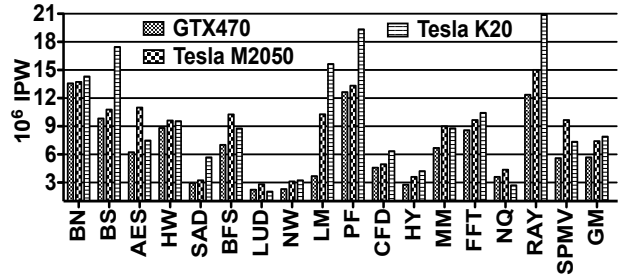

Fig. 1. Energy efficiency of throughput workloads.

In Figure 1, we show the million instruction-per-watt (IPW) of various non-concurrent throughput workloads executed on different generations of Nvidia GPUs. Average performance-per-watt is significantly less than the theoretical peak of the respective GPUs (GTX470: 294, Tesla M2050: 599, Tesla K20: 1909 $10^6$ IPW). These GPUs are capable of concurrent execution of multiple kernels. Note that, significant overlap of throughput kernels increases hardware utilization, performance, and offers regulated power delivery opportunity across kernels to improve energy efficiency. Although multi-kernel benchmarks aim to enhance the use of processing elements inside the throughput core, existing single kernel compute bound workloads can exhaust the GPU core. Contrarily, in memory bound workloads, the processing elements often have to wait for slow off-chip data that leads to under-utilization of the throughput core. Intuitively, concurrently running compute and memory bound workloads can improve the utilization of the GPU. When the memory-bound workload exhausts its entire thread pool and waits for memory access to finish, the simultaneously running compute-bound kernel begins execution. It will not only improve utilization and performance but also will raise the power efficiency.

### 2.2 Concurrency in Throughput Architecture

In throughput architecture, concurrency exists at different levels: application, kernel, task, thread, and data (See Figure 2). Various applications create different *contexts* [26] (equivalent to *process* in CPU). In each context, different *streams* [26] run a sequence of operations dedicated to solving a single problem. Host CPU schedules *stream* to the throughput processor. Overlap across *streams* refers to as job-level concurrency. Each stream often has a set of tasks, which are different in types (CUDA *kernels*, bulk memory operations) and computation. Various tasks can be performed simultaneously to improve throughput processor utilization and energy efficiency. Task-level parallelism often requires simultaneously executing workloads to overlap memory transfers from one workload with computation from another. Having adequate hardware resources enables simultaneous execution of intra-/inter-kernel thread-blocks. Intra-block threads run in parallel in SIMD hardware, which refers to as thread level concurrency. Vector instructions within a throughput thread exploit data level parallelism. Thread and data level concurrencies are beyond the scope of this work.

Concurrency is limited by several factors at various levels, [25] reports such findings. Excluding resource limitation, long memory transfers and long-running kernels can both restrict small kernels and memory transfers from being scheduled due to dependency. Suboptimal stream scheduling results in degraded performance. In Nvidia imple-mentation of throughput API, memory operations (*cudaMalloc/cudaFree/cudaMemset*) serialize execution. Moreover, specific throughput architecture generations (Nvidia Fermi) often restrict inter-kernel/inter-memory-transfer concurrency due to the false assumption of inter-dependency.

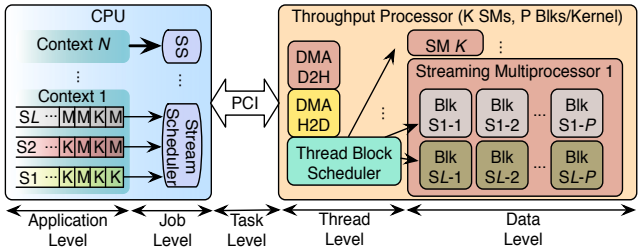

Fig. 2. Concurrency in throughput architecture.

## 3 CHARACTERIZATION STEPS

In this section, we provide methodology details. Figure 3 depicts the flow of operations for multi-kernel workload generation. Following subsections explain the process in detail.

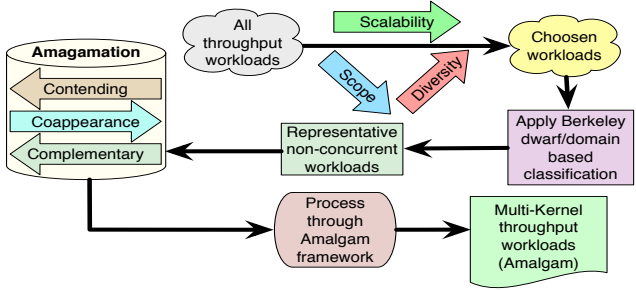

Fig. 3. Characterization methodology in steps.

### 3.1 Throughput Benchmark Selection

We have used Berkeley Dwarves [27] based systematic approach for throughput workload selection. Table 1 shows various Berkeley dwarves covered in this work. To choose representative workloads that include the dwarves, we have explored Nvidia GPU computing SDK workloads [14], Rodinia workloads [13], Parboil workloads [12], and

### TABLE 1
### BERKLEY DWARVES COVERED

| | Berkeley Dwarves | Example |
|---|---|---|
| Preliminary | 1. Structured Grid | Sum of absolute difference (MPEG2) |
| | 2. Unstructured Grid | Computational fluid dynamics solver |
| | 3. Dense Linear Algebra | LU factorization |
| | 4. Sparse Linear Algebra | Matrix multiplication |
| | 5. Particle Method | Particle potential calculation in 3D space with multiple particles |
| | 6. Monte Carlo Method | Option pricing algorithm |
| | 7. Spectral Method | Fourier transform |
| Auxiliary | 8. Combinational Logic | Encryption algorithm (AES) |
| | 9. Graph Algorithm | Breadth first search |
| | 10. Dynamic Programming | Sequence alignment |
| | 11. Back-track | N-Queen solver |
| | 12. Graphical Models | Ray tracing |
| | 13. Sorting | Bucket sort, quick sort and merge sort |

several third party benchmarks [11], [28]. Workload selection process scrutinized the application purview (data centers, mobile, desktop, embedded) of the workload, characteristic diversity of the benchmark based on [12], [13], [29] and scaling (with growing load) capability of workloads in scalable emerging systems. Application scope ensures broader impact than the today's state-of-the-art. The characteristic diversity guarantees architecture exploration capability. The scalability captures adaption ability of the workloads with larger input/system in the future. Moreover, the chosen workload set simultaneously covers various application domains [30] such as *high performance computing, finance, image processing, audio processing, video processing, health applications, graphical models, database, browser, general algorithms* (sorting, searching, grid traversal etc.) and all the Berkeley Dwarves [27].

### 3.2 The Methodology

In this section, we propose a multi-kernel throughput workload generation framework.

#### 3.2.1 Performance-power Co-characterization of Workloads

It comprises of three steps. To start with (step 1), microarchitecture agnostic workload behaviors [29] and microarchitecture dependent power-performance characteristics categorize benchmarks in Table 2. Tables 3 and 4 list the characteristics. The set of microarchitecture agnostic metrics unleashes two types of behaviors. One is the intrinsic behavior using generic workload characteristics such as dynamic instruction count, memory / branch / atomic / shared-memory instruction count, etc. The other is the throughput workload specific traits using per-thread register usage, data transfer in between host and device, control flow divergence, memory access locality, thread-batch efficiency, etc. On the contrary, power-performance metrics express power and performance dissimilarities to help the co-characterization process. Precisely, power, energy, and temperature depict energy consumption aspect of the workloads. IPC indicates performance. The communication overhead encapsulates performance degradation due to excessive host to device interaction. Finally, IPW/EDP captures co-optimization characteristics. Note that the tables have inter-dependent parameters.

In the next step (step 2), workloads from Table 2 are executed on real Nvidia hardware (see Section 4) such as Tesla

M2050, Tesla K20X, and GTX470. Using Nvidia Nsight

### TABLE 2
#### Throughput Workload Synopsis

| Bench | Domain (Dwarf) | Problem Size |
|---|---|---|
| Breadth First Search (BFS) [13] | General algorithms (Graph algorithm) | Graph with 1 million nodes |
| Sum of Absolute Difference (SAD) [12] | Video processing (Structured grid) | 1920.1072 frame sequence |
| LU Decomposition (LUD) [13] | High performance computing (Dense linear algebra) | Matrix size 2048.2048 |
| Matrix Multiplication (MM) [14] | High performance computing (Sparse linear algebra) | Matrix size 5120.10240 |
| Black Scholes (BS) [14] | Finance (Monte Carlo method) | 4M options in 512 iterations |
| Binomial Options (BN) [14] | Finance (Monte Carlo method) | 4M options |
| Path Finder (PF) [13] | General algorithms (Dynamic programming) | 2D space of size 800000,800 |
| 2D Convolution using FFT (FFT) [14] | Audio processing (Spectral method) | 2D convolution using FFT of size 2048.2048 |
| Ray Trace (RAY) [14] | Computer graphics (Graphical models) | Rendering image of size 2048.2048 |
| Computational Fluid Dynamics Solver (CFD) [13] | High performance computing (Unstructured grid) | 97K elements |
| Sparse-Matrix Dense-Vector Multiplication (SPMV) [12] | High performance computing/Image processing (Sparse linear algebra) | Matrix size 146689,146689 |
| Heart Wall (HW) [13] | Health application (Structured grid) | Heart tracking in 50 frames of size 609.590 |
| Hybrid Sort (HY) [13] | General algorithms (Sorting) | List of $2^{20}$ elements |
| Needleman-Wunsch (NW) [13] | Health application (Dynamic programming) | Sequence size of 16384 |
| N-Queen Solver (NQ) [11] | High performance computing (Back-track) | Chess board of size 16.16 |
| Advanced Encryption Standard (AES) [28] | High performance computing/database (Combinational logic) | 128 bit encryption of 256KB image |
| Lava MD2 (LM) [13] | High performance computing (Particle method) | 0.8M particles in 8K boxes |

### TABLE 3
#### Throughput Workload Characteristics

| Characteristics | Synposys |
|---|---|
| Registers/Thread | Number of registers used per thread |
| Shared Memory | Amount of shared memory used per thread |
| Branch Efficiency | Percentage of non-divergent branches |
| Thread Batch Efficiency | Percentage of non-divergent thread batches |
| Kernel Count | Total number of kernels |
| Thread Count | Total number of threads launched |
| Dynamic Instructions | Dynamic instructions count across all kenels |
| Local Memory Inst. | Local memory load-store count |
| Global Memory Inst. | Global memory load-store count |
| Shared Memory Inst. | Shared memory load-store count |
| Branch Instructions | Total branch instructions count |
| Divergent Branches | Total divergent branch instructions count |
| Atomic Instructions | Total atomic instructions count |
| Device to host Transfer | Device to host data transfer in bytes |
| Host to Device Transfer | Host to device data transfer in bytes |
| Off-chip Efficiency | Percentage off-chip row access locality |

Eclipse [31] hardware profiler, we have collected all the metrics of Table 3 and 4. We have performed two separate Principal Component Analysis (PCA) and clustering (hierarchical, K-means) analyses based on the two tables. Such analysis reveals similarity and dissimilarity information across the benchmarks and assists in selecting representative kernels. Figures 4 and 5 depict the throughput workload PCA (PC 1-5, 73% variance) plot and hierarchical clustering. The PC1-PC2 (42% variance) plot in Figure 4 shows that BS, RAY, NQ, SPMV, and PF have distinctive individual properties (placed in different corners). Based on workload scattering in various PC domains, we confirm that microarchitecture-independent characteristics successfully retains workload diversity. Figures 6, 7, and 8 show the results of power analysis based workload clustering on Tesla M2050, K20, and GTX470. Across GPUs with different power efficiency, power behavior based clustering changes significantly.

### TABLE 4
#### Power Performance Behavioral Metrics

| Characteristics | Synposys |
|---|---|
| Average Power ($P_A$) | Average power across various kernels |
| Peak Power ($P_P$) | Maximum power across various kernels |
| Total Energy (E) | Total energy consumption for the workload |
| Instruction-per-Watt | Average power per instruction |
| Energy-Delay-Product | Energy multiplied by execution time |
| Instruction-per-Cycle | Average instructions executed per cycle |
| Instruction-per-Second | Average instructions executed per second |
| Execution Duration | Kernel execution time |
| Comm. Overhead | Number of memory transfer commands |
| Maximum Temperature | Max temp. for fixed initial temperature |
| Average Power | Average power across various kernels |
| Peak Power | Maximum power across various kernels |
| Total Energy | Total energy consumption for the workload |
| Instruction-per-Watt | Average power per instruction |
| Energy-Delay-Product | Energy multiplied by execution time |
| Instruction-per-Cycle | Average instructions executed per cycle |

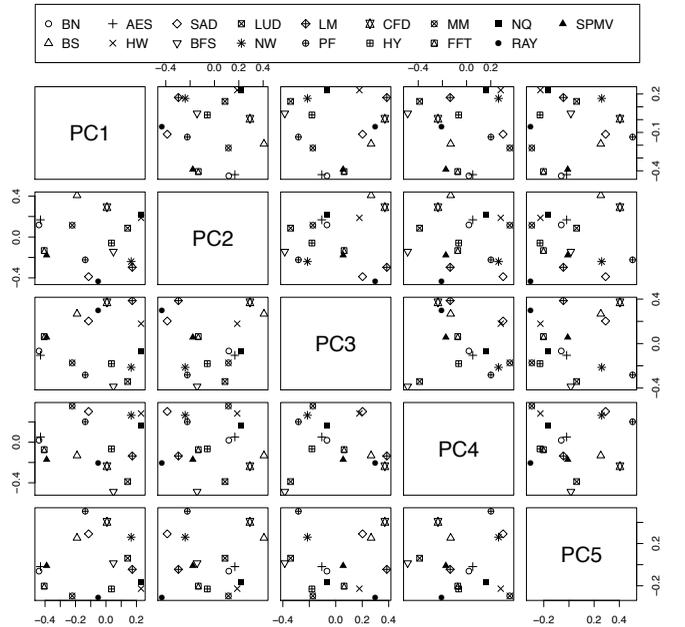

Fig. 4. PCA based on microarchitecture independent characteristics (variance 73%)

Finally, (step 3) to choose a set of representative multi-kernel throughput workloads, we assign a *relation score* to each benchmark and create a workload database (See Figure 9). Each benchmark in a cluster receives a score. There are multiple such clusters generated from the power and characteristics analysis. To avoid clustering artifact, we performed *hierarchical* and *kmeans* clustering simultaneously on all data. Note that; the final score also depends on the architecture agnostic workload characteristics and individual power-performance characteristics. Since workload characteristics define the execution pattern, we assign greater weight to it. We have used cluster ensemble analysis [32] on characteristic and power clusters for each GPU. Output clusters are further fed into another *cluster ensemble analysis* to obtain consensus cluster for a given technique (kmeans/hierarchical) (See Table 5). Inter-technique cluster ensemble analysis provides the final set of clusters mentioned in Table 5 and individual benchmark scores in the workload database. In the next stage, we use workload database to generate parallel benchmarks based on the integration strategy.

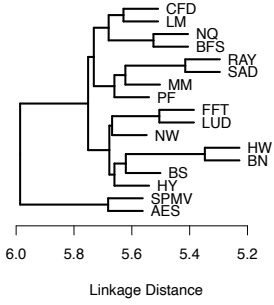

Fig. 6. Dendrogram for microarchitecture independent properties.

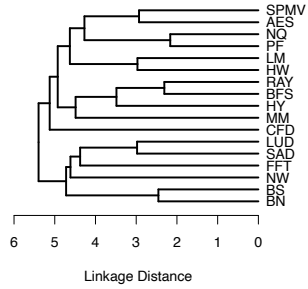

Fig. 7. Dendrogram for power-performance on Tesla M2050.

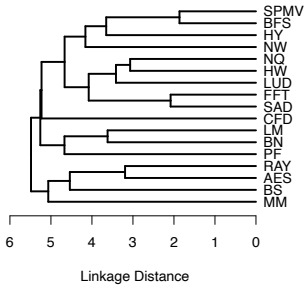

Fig. 8. Dendrogram for power-performance on Tesla K20.

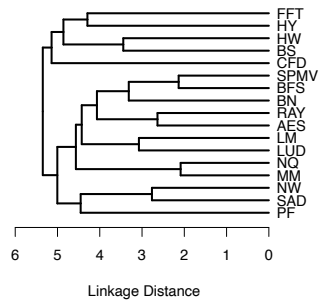

Fig. 9. Dendrogram for power-performance on GTX470.

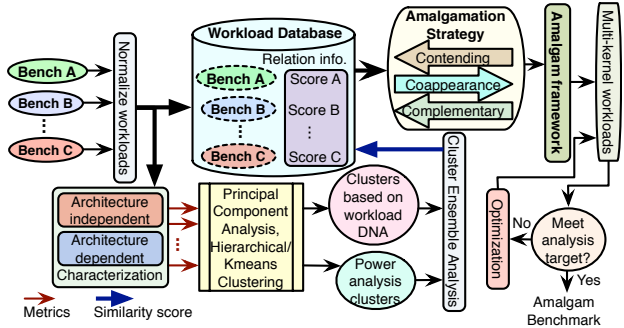

Fig. 5. Operation flow in flexible benchmark generation.

### 3.2.2 Integration Strategy

After exploring single kernel throughput workload space, we propose a novel single kernel workload integration strategy that retains three main features of computer architecture workloads: suitable representation, module-level hardware stress generation ability, and complete hardware coverage. To achieve this, we segregate the multi-kernel workloads in the three top categories: contending, complementary and co-appearance based integration.

An architect interested in exploring the effectiveness of a specific hardware design optimization needs resource contention in the module. For example, to optimize the control-flow-divergence mitigation module, an architect can regulate the stress on the device by appropriately launching multiple divergent kernels simultaneously. Based on the analysis, FFT and LUD kernels behave similarly regarding execution pattern, and power-performance behavior thus represents a contending workload.

To see the overall impact of proposed throughput architecture, architects often require workloads that cover the design gamut. Such workloads reveal the interaction of various hardware modules and ability to operate in unison. Unlike traditional workloads, complementary integration based multi-kernel workloads promise regulated and wider hardware block coverage.

Co-appearance based integration reflects the real life scenario when distinct kernels often execute simultaneously. For example, in the multimedia stream decoding, FFT (audio) and SAD (video) are often performed concurrently.

TABLE 5
THROUGHPUT WORKLOAD CLUSTERS

| Kmeans | Hierarchical | Consensus |
|---|---|---|
| BN, CFD | BN, BS, HW | BN, HW |
| NW, LM | AES, SPMV | BS, HY |
| AES, RAY, BFS | SAD, MM | AES, SPMV |
| BS, HY | BFS, NQ | MM, NQ |
| NQ, MM | LUD, FFT | BFS, RAY |
| SAD, PF | NW, PF | FFT, LUD |
| LUD, FFT | LM, CFD | NW, PF |
| SPMV | HY | LM, CFD |
| HW | RAY | SAD |

### 3.2.3 New Benchmark Suite and the Framework

Based on integration strategy and Table 5 workload classification, we propose the new multi-kernel benchmark suite in Table 6. Benchmarks from different far clusters (MM, BFS, BS) clubs in a single multi-kernel workload, and it represents significantly different behavior. As a complementary workload, it has distinct execution pattern, resource utilization, and power behavior. Workloads from the same cluster are similar in nature with similar stress pattern. FFT and LUD produce one such mix. For set size 3 and 4, we club benchmarks from close clusters. SAD/FFT, BN/BS and BFS/HY represent few common execution scenarios in multimedia processing and HPC datacenters. Various real world applications require frequent searching and sorting operations. Breadth-first-search and hybridsort (BFS/HY) represent such a case. In finance, different option-pricing algorithms are used to predict stock price. Using multiple such algorithms simultaneously is also a common use case. BlackScholes and Binomial-Options (BS/BN) represent commonly used finance applications. There are many such possible cases and due to time and space limitation we have chosen the most common cases.

**TABLE 6**
NEW BENCHMARK SUITE

| Complimentary | Contending | Co-appearance |
|---|---|---|
| (BFS, PF), (NW, BS), (NQ, LM), (SPMV, RAY), (SAD, LUD, BN) | (MM, LUD), (LM, CFD), (SPMV, AES), (AES, NW, HW) | (SAD, FFT), (BN, BS), (BFS, HY) |

Multi-kernel benchmark generation process is automated and flexible. The Benchmark generation framework takes several input parameters and single kernel benchmark source code as input to produce the output benchmark. Input parameters include benchmark resource usage information, throughput API launch sequence (breadth-first: H2D-H2D-K-K-D2H-D2H, depth-first: H2D-K-D2H-H2D-K-D2H, custom, etc.), memory transfer slicing (large memory transfers are subdivided into smaller pieces to avoid delays), etc. Note, H is the host CPU, D is the GPU and K is the kernel. This output multi-kernel benchmark may or may not meet the power-performance goals desired by the end-user. Therefore, we further optimize the benchmark to fine-tune it. The optimizations are often workload integration specific and require kernel level code optimization.

### 3.2.3.1 Engineering Kernel Concurrency

Simultaneous execution of kernels is achieved by changing the resource allocation of the throughput workload. As mentioned in [21, 25], we modify the granularity of work-item (in Nvidia terms block) collection (in Nvidia terms grid) size. To avoid losing inherent performance optimization capability, to avoid violating resource (memory usage, synchronization, etc.) usage restriction and to perform additional engineering within a work-item, we consider modifying the work-collection size only. Kernels are launched from different host threads to maximize the execution overlap of different work-collections from various kernels. Pre-Kepler architecture benefits from this technique due to single hardware launch stream shared among multiple logical streams. However, in several instances, we found that thread based work-collection launch is inadequate to achieve concurrency. For example, NQ_LM or BFS_PF benchmarks have memory-kernel-memory sequence in a loop. Memory and Kernel operations have varied execution latency. In pre-Kepler Nvidia architectures, a memory operation is blocked due to signals between sequentially issued kernels. Hence, we align throughput subroutine invocations from different kernels by changing the call sequence. Such optimization achieves memory and kernel overlap in BSF_PF in Fermi GPUs too.

### 3.3 Multi-kernel Workload Characteristics

Unlike single kernel throughput workloads, new benchmarks possess different performance and power traits. Excluding *throughput, processor utilization* (core occupancy), *energy efficiency* (sum of all single kernel energy consumptions divided by multi-kernel energy), and *power efficiency*

(instruction-per-watt), we define the following metrics to capture natural multi-kernel workload power-performance behaviors.

### 3.3.1 Impact on Power due to Overlap (IPO)

Inter-benchmark kernel and memory transfer overlap are expected to change the intrinsic power characteristics of the benchmarks. Using IPO, we capture the percentage of power reduction in multi-kernel workload compared to single kernel workload due to overlapped execution. It represents average and peak power saving due to overlap. Peak power is considered due to its substantial impact on the cooling infrastructure of power hungry throughput processors.

### 3.3.2 Impact on Energy due to Overlap (IEO)

Percentage reduction in overall energy consumption due to overlapped execution is collected using this metric. Unlike *energy efficiency*, this metric shows the benefit of overlap in energy saving. It is a ratio of energy saving to overlap duration.

### 3.3.3 Impact on Power-performance Co-Characterization (IC)

Availability of throughput processors in HPC datacenters influence job scheduling. Even for non-blocking execution scenarios, the host CPU will wait for throughput processor's computed data to become available. It increases overall CPU busy time and idle power. Simultaneous optimization of energy consumption and performance will capture best of both worlds to improve the overall responsiveness and the power efficiency. Hence, using *energy-delay-product*, we show that throughput processor can be more responsive to multi-kernel execution.

**TABLE 7**
HARDWARE TESTBED

| | GTX 470 | M2050 | K20 |
|---|---|---|---|
| Compute Capability | 2.0 | 2.0 | 3.5 |
| Core Frequency | 1.12 GHz | 1.15 GHz | 0.7 GHz |
| GFlops (SP) | 633.6 | 1288 | 4106 |
| SM Architecture | Fermi | Fermi | Kepler |
| Main Memory GDDR5 | 1.3 GB | 3 GB | 5 GB |
| Power (TDP) | 215 W | 225 W | 225 W |
| Use scope | Graphics | HPC | HPC |
| Data Acquisition Sys | Lab View NI PCI-6221 | | |
| Current Sensor | HCS-20-10-AP | | |
| Host CPU | Intel i7 2600S, 4-core, TDP (65 W) | | |
| Motherboard | GigaByte GA-H77M-D3H | | |
| CPU Main Memory | 32 GB | | |
| OS | Ubuntu 12.04 | | |
| Program Model | CUDA 5.0 | | |

## 4 EXPERIMENTAL SETUP

Benchmark generation uses real GPUs as throughput processors for experimentation. Table 2 lists the single kernel benchmarks.

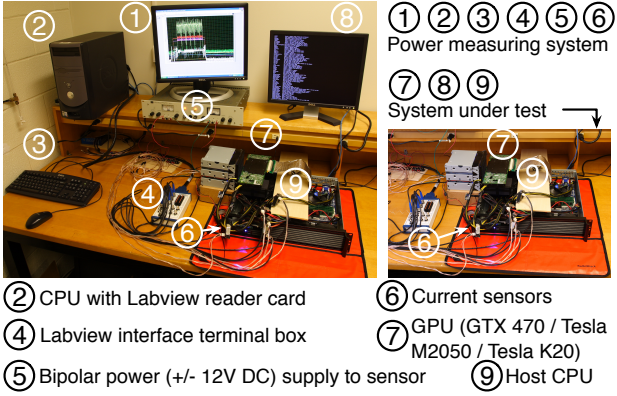

① ② ③ ④ ⑤ ⑥
Power measuring system

⑦ ⑧ ⑨
System under test

② CPU with Labview reader card

④ Labview interface terminal box

⑤ Bipolar power (+/- 12V DC) supply to sensor

⑥ Current sensors

⑦ GPU (GTX 470 / Tesla M2050 / Tesla K20)

⑨ Host CPU

Fig. 10. The testbed.

## 4.1 System Architecture
### 4.1.1 Hardware

Figure 10 shows a pictorial view of the test-bed. Table 7 lists our hardware infrastructure details. We have used three different generations (GTX470: gaming, M2050: Fermi HPC GPU, K20: Kepler HPC GPU) of Nvidia GPUs as throughput processors. Microarchitecture and use scope variation ensure robust workload analysis. Instead of simulation-based approach, we use real GPU based profiling data to guarantee a widely applicable result. The experimental platform is configured with an Intel i7 CPU and a GPU installed in the PCI slot through a riser card for easy access to power supply lines. The power supply feeds the GPU via 12V and motherboard via 12V, 5V, and 3.3V supply lines. Besides, the motherboard feeds the GPU with 12V and 3.3V PCIe power lines via the riser card. Instead of measuring the overall server power using an external meter, we opt for a detailed power profile through simultaneous current measurement on individual power lines via Hall Effect current sensors. The sensors produce a voltage proportional to current passing through the clamped wire. The current sensors have an output voltage range of +/- 4V, an accuracy of +/- 1%, and a response time of less than 3 microseconds. We have validated the accuracy of the sensors through calibration against precision current circuits and resistances. A LabView virtual instrument controlled a data acquisition card NI PCI-6221 to collect current profile data from current sensors simultaneously. The sensors were sampled at 250 kS/s. To verify the accuracy of collected data, we validated the profiles with a high-performance multifunction data acquisition card (PCI-6110) with a sampling rate of 5 MS/s. Power and energy consumptions were computed from the collected raw profile data.

### 4.1.2 Software

Throughput processor performance and intrinsic workload characteristics are obtained using Nvidia Nsight Eclipse Edition [31] software profiler that uses *nvprof* profiler. We have used CUDA programming model 5.0 within Linux. New workload generator is written in Python. Figure 11 shows power profiles for (NW, BS) three GPUs running the benchmark. As evident, there are variations in the power drawn by input lines for each GPU. In particular,

K20 draws significantly less power than GTX and Tesla GPUs. This change is manifested in our power-performance analysis discussed in Section 5.

## 4.2 Evaluation Approach

To ensure diversity of workloads, we have performed multi-kernel benchmark evaluation. Since the new workload behavior is significantly different from single kernel workloads, a set-to-set comparison using performance and power metrics might not be representative. Hence, we use PCA/clustering based analysis to observe the intra-suite diversity. Average power and performance of the single kernel benchmarks and the multi-kernel benchmarks are also compared to the whole suite to see the impact of multi-kernel benchmark on the power and the performance. Single kernel benchmarks are executed as many times as they are used in the whole suite. For example, BFS is executed 4 times.

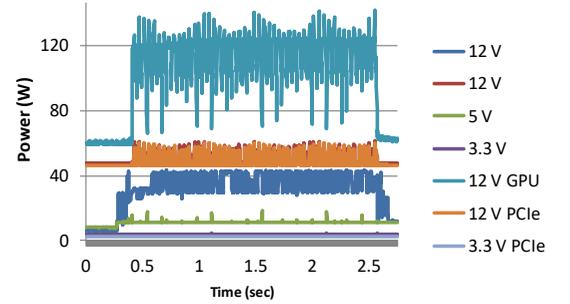

(a) GTX 470

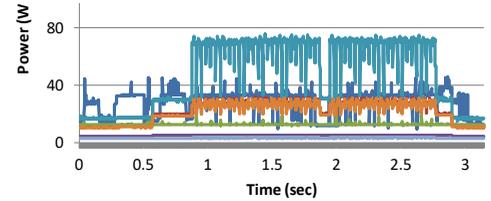

(b) K20

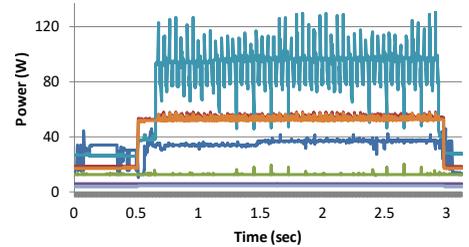

(c) Tesla M2050

Fig. 11. The power profiles.

## 5 RESULTS AND ANALYSIS

In this section, we analyze new workload characteristics in terms of diversity and power-performance behaviors.

### 5.1 Workload Diversity
### 5.1.1 Intrinsic Behavior Versatility

Figures 12 and 15 show the principle component analysis (73% cumulative variance) and hierarchical clustering of all concurrent and sequential kernel workloads respec-tively. The PCA confirms workload space diversity; the single kernel and the concurrent kernel workloads are spread across the entire chart. In particular, PC1 (30% variance) - PC2 (15% variance) plot, which retains 45% of the information, indicates that benchmarks such as LUD, NW_BS, BN, LM_CFD, SPMV_RAY position themselves at various corners of the chart and center. Figure 13 reinforces our claim that sequential and concurrent kernel workloads possess distinct intrinsic characteristics. In most of the cases, concurrent kernels have considerable linkage distance from the sequential kernels. Interestingly, often one or more kernels in the new workload show characteristic dominance within the benchmarks. For example, in LUD, MM_LUD, and SAD_LUD_BN, the dominant behavior of LUD keeps them in proximity. A similar trend is visible in BFS, BFS_PF, and BFS_HY. On the contrary, FFT_LUD is not close to other LUD workloads. Here distinct behavior of FFT subdues the LUD dominance. In essence, the insights are: *Quantitative evidence suggests that sequential and concurrent kernels are indeed dissimilar in nature. Intrinsic workload characteristics of sequential kernels dominate or subdue co-existing kernels in concurrent workload behavior.*

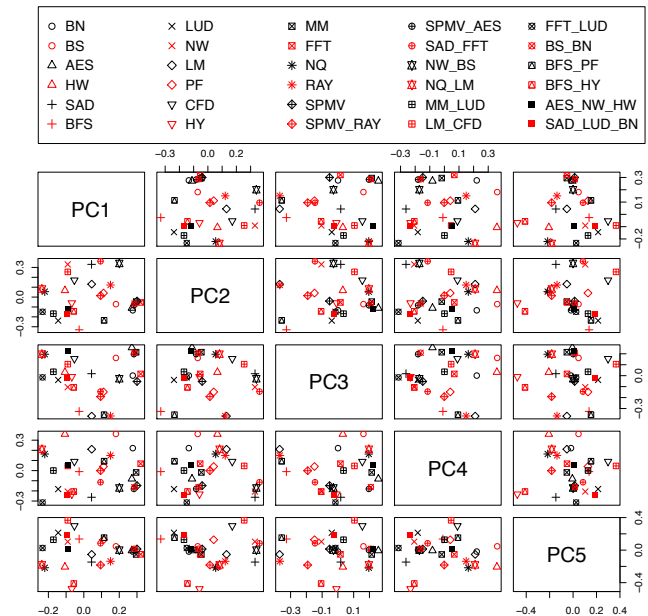

Fig. 12. PCA of the all benchmarks based on microarchitecture independent characteristics (72% variance).

### 5.1.2 Power-performance Diversity

Figures 13, 14 and 16 show power-performance co-characterization of concurrent kernel workloads. The PCA is performed based on first 3 PCs. In Tesla M2050, Tesla K20 and GTX470, the 3 PCs retain 95%, 93% and 94% cumulative variance respectively. Clustering trends in three generations of throughput architecture are distinct. For example, at high linkage distance of 3 in all the dendrograms, Tesla M2050, GTX470 and Tesla K20 have no common clusters. As claimed by Nvidia, GTX470, M2050 and K20 GPUs show prominently different power-performance traits. Both being Fermi architectures, GTX470 and Tesla M2050 have different power efficiency due to power overhead of graphics capability of GTX470. Order of magnitude power

efficiency improvement of K20 is clearly visible from M2050 and K20 clusters. Interestingly, concurrent kernels (SPMV_RAY, SPMV_AES) with a common sequential kernel (SPMV) do not cluster together in any dendrogram. We observe similar power-performance trend among several concurrent workloads in different GPUs: (NW_BS - SPMV_AES, BFS_PF – FFT_LUD, BS_BN – SAD_FFT) in GTX470, (SAD_FFT - SPMV_RAY, SAD_LUD_BN – NW_BS) in M2050 and (BFS_HY – NW_BS, BFS_PF – SPMV_RAY – SAD_LUD_BN) in K20. Unlike, intrinsic workload characteristics, individual workload dominance or subduing trend is absent in power-performance behavior of concurrent kernels. Certainly, proposed throughput power-performance behaviors are microarchitecture-governed.

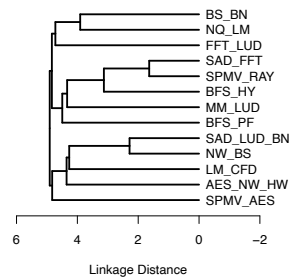

Fig. 13. Dendrogram using power-performance characteristics Tesla K20.

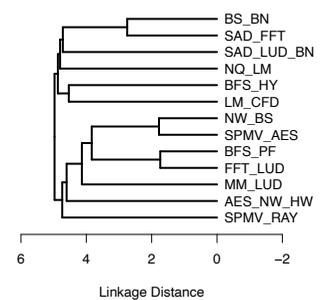

Fig. 14. Dendrogram based on Power-Performance Characteristics GTX470.

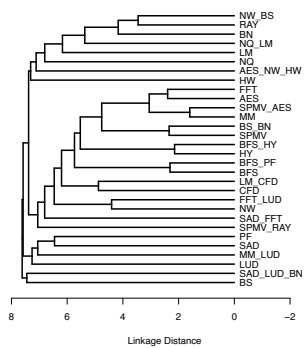

Fig. 15. Dendrogram based on microarchitecture independent characteristics of all benchmarks.

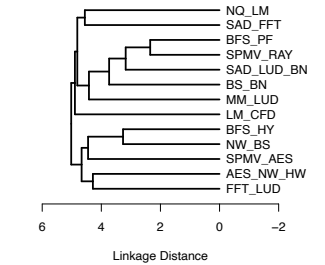

Fig. 16. Dendrogram based on power-performance characteristics of Tesla M2050.

## 5.2 Impact of Concurrency on Energy and Power

Figures 17, 18 and 19 summarize the impact of kernel execution concurrency on power dissipation and energy consumption of new workloads. Concurrency represents the percentage of time the kernels are executed simultaneously. Based on the subroutine invocation pattern and execution duration, kernels achieve varied concurrency. It was as small as 5% in NQ_LM and up to 100% in NW_BS. Extended kernel execution period and limited resource utilization by BS make room for BN to execute simultaneously. Interestingly, invocation latency of work-collection also regulates the amount of possible concurrency. Due to single hardware stream in Tesla M2050 and GTX470, work-

collections in HY overlap with the memory transfer operation of BFS in BFS_HY benchmark where H2D-K-D2H executes in a loop. The action sequence is $H2D_1$-($K_1$-$H2D_2$)-($K_2$-$D2H_1$)-$D2H_2$. On the contrary, the Kepler architecture executes kernels simultaneously such as *(H2D-K)-(K-K)-(H2D-K)-(K-K)-(H2D-K)*. Actions in brackets execute simultaneously. As expected, Kepler K20 achieves maximum average concurrency of 55%, and Fermi architectures reach average concurrency of 46%, 47% respectively for GTX470 and Tesla M2050. *Note, both work-collection size and launch time control the amount of attainable concurrency.*

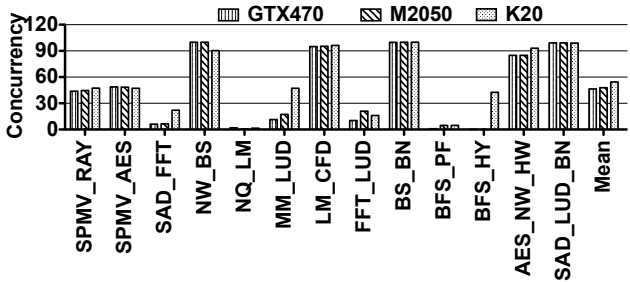

Fig. 17. Observed concurrency in kernel execution.

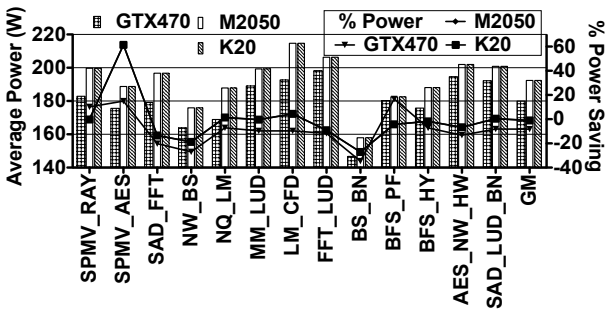

Fig. 18. Average power and percentage power saving in concurrent kernels.

Average kernel power dissipation of parallel workloads is different from the sequential counterpart. Intuitively, concurrent execution suggests greater hardware involvement in computation, thus requiring more power. However, from thread-batch execution perspective, throughput cores will often have more active thread-batches to schedule while one or more thread-batches are stalled (inactive) due to off-chip memory accesses.

Note that; total active thread-batch count is unchanged. It is the heterogeneity within the resident thread-batch pool that regulates utilization and power dissipation. Power saving in concurrent kernels varies from +61% (high hardware utilization) in SPMV_AES to -34% (reduced hardware utilization) in BS_BN. SPMV_AES has twice the occupancy of BS_BN. Moreover, SPMV and AES have various kernels, which increases heterogeneity and usage. On the contrary, BS and BN show similar kernel invocation pattern. For most kernels, M2050 and K20 power dissipations are close. GTX470 dissipates around 10W less average power. Although it is counter-intuitive, energy consumption profile in Figure 19 elucidates the contradiction. Most of the workloads consume more energy in GTX470 than in other GPUs. On average, GTX470, M2050, and K20 save 32%, 26%, and 33% energy. Higher clock frequency in GTX470 expedites processing and reduces energy. Power efficiency optimization in K20 improves the overall energy consumption of the GPU. Figure 17, 18 and 19 explains the IPO and IEO metrics. *Contemporary energy efficient throughput architectures consume comparatively more power, but reduced execution time saves overall energy.* It is verified that kernel execution concurrency attributes to overall energy saving.

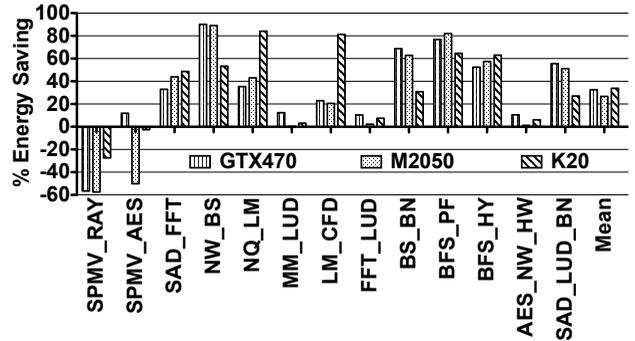

Fig. 19. Energy saving in concurrent kernels.

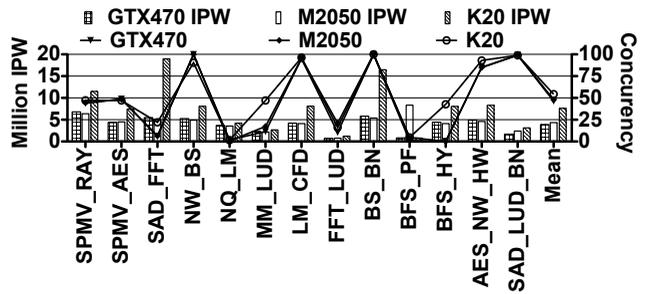

Fig. 20. Impact of concurrency on power efficiency.

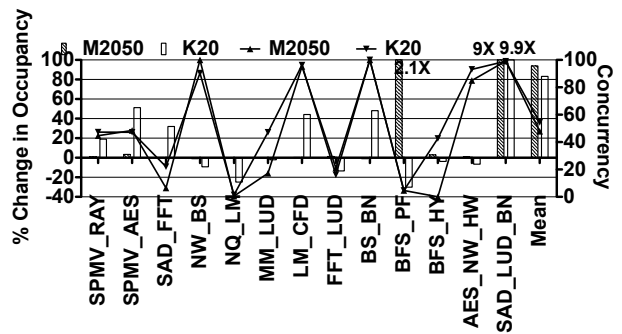

Fig. 21. Correlation between kernel concurrency and occupancy.

## 5.3 Power Efficiency and Occupancy Analysis

Figures 20 and 21 provide the work done for unit energy consumed and occupancy variation in different generations of GPUs. On average, concurrent workloads achieve 4, 5 and 7 Million IPW respectively for GTX470, M2050 and K20 GPUs. This trend is intact for most of the workloads. Excluding SAD_FFT, LM_CFD, AES_NW_HW, and SAD_LUD_BN, all the other benchmarks show a linear correlation between concurrency and power efficiency. In LM_CFD, higher concurrency increases power due to increased utilization, which finally attributes to decreased IPW. The SAD_FFT suffers from lower concurrency and low IPW in M2050 and GTX470. However, 25% improve-ment in concurrency for K20 is attributed to higher utilization and power efficiency. The LUD has more inferior IPW due to lower hardware utilization (63%). All the benchmarks with LUD suffer from lower IPW. Sequentially running AES, NW and HW kernels have lower utilization; this reduces power efficiency of all the multi-kernel workloads that include them. Concurrency alone does not guarantee higher power efficiency; together concurrency, execution pattern, and resource contention trait often dictate improved power-performance co-optimization.

Comparative analysis of occupancy (% core utilization) in Figure 21 reveals that, on average, M2050 and K20 achieve 91% and 83% more occupancy compared to GTX470. However, benchmarks with lower concurrency (NQ_LM, SAD_FFT, FFT_LUD) often suffer from lower average occupancy. Our experiments confirm that concurrently running kernels improves processor resource utilization.

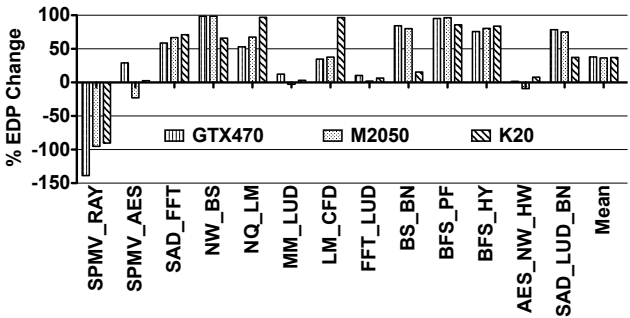

Fig. 22. % change in EDP compared to sequential kernel execution.

### 5.4 Power Performance Co-characterization

Using energy-delay-product (EDP) in Figure 22, we explain the impact of power-performance co-optimization metric (IC) in concurrent workloads. On average, compared to sequential kernel launch, parallel execution improves EDP by 37%, 36% and 37% in GTX470, M2050, and K20 respectively. Improved EDP also guarantees enhanced responsiveness at lower energy budget. However, benchmarks such as SPMV_RAY, SPMV_AES, and AES_NW_HW show degraded EDP. These benchmarks suffer from higher energy consumption due to increased average power (SPMV_AES) and longer execution time (SPMV_RAY, AES_NW_HW). Due to inherent characteristics and power profile, most of the benchmarks follow either increasing or decreasing EDP trend in GTX470, M2050, and K20. Evidently, throughput kernel concurrency improves power-performance co-optimization and responsiveness of the GPUs.

## 6 RELATED WORK

We distinguish our work from three major aspects:

### 6.1 Workload Characterization

In [33], Saavedra et al. have demonstrated workload characteristics based benchmark characterization. Eeckhout et al. [34, 35] extended the work by exploring the correlation between the characteristics to remove redundant workloads using PCA and clustering techniques. CPU2000 and SPEC2006 suites are inspected concerning stress behavior on different architectures in [36] and [37] respectively. In [38], Hughes et al. have characterized the transaction memory workloads. In [52], Esmaeilzadeh et al. presented quantitative power and performance measurement at the chip level across hardware generations using single threaded and multithreaded, native and managed workloads. Cao et al. in [53] has explored asymmetric multi processor based software/hardware co-design for big and small cores and the system software, and in particular VM services, for optimized power-performance results. Unlike all above, we have characterized multi-kernel throughput workloads. We have extended the exploration by including microarchitecture dependent metrics such as power related metrics.

In [29], Goswami et al. have characterized several GPGPU workloads using various microarchitecture-independent parameters. The exploration is based on the profiling data obtained from GPGPU-Sim [39] throughput processor simulator. Unlike [29], we provide more robust results by profiling the workloads on real hardware and by using hardwired power analysis framework. In [10, 13, 40], Che et al., Kerr et al. and Wenbin et al. have characterized Rodinia benchmark suite, Nvidia CUDA SDK, Parboil benchmark suite and Mars benchmarks respectively. Che et al. in [41] have further compared Rodinia throughput workloads with contemporary CMP benchmarks suites such as Parsec [42]. Burtscher et al. [9] have analyzed the control-flow and the memory access diversity of graph algorithm on throughput architecture. Excluding [13, 41], others did not consider systematic statistical analysis based workload characterization. In [13], authors used MICA framework to analyze the workloads. However, none of these works include power characterization or multi-kernel throughput workload analysis.

### 6.2 Throughput Architecture Power Analysis and Models

In [43], Nagasaka et al. have done single kernel throughput workload power analysis using GTX 285 GPU. Using linear regression-based power modeling, they have analyzed the correlation between real hardware power measurement and modeled power data. Rofouei et al. in [44] have demonstrated novel platform for runtime energy dissipation collection framework. Huang et al. have explored the power profile of heterogeneous architectures in [45]. Ren et al. used a similar approach in performing a detailed study by varying throughput workload resource allocation and architectural parameters in [46]. In [47], authors have proposed a throughput power model based on a tree based random forest methods. Goswami et al. in [48] and Leng et al. in [49] have proposed throughput architecture power model. Our exploration reveals the multi-kernel throughput workload power behavior based on real GPU based power data.

### 6.3 Multi-kernel based Throughput Architecture and Resource Sharing

To the best of our knowledge, this is first work that explores the power behavior of multi-kernel throughput workloads.

Pre-Fermi (kernel concurrency not supported) GPU concurrency is analyzed by Guevara *et al.* in [50] by compile time kernel fusion. OpenCL based runtime *KernelMerge* framework is proposed by Gregg *et al.* in [24]. It explores several thread-block level scheduling schemes. Wang *et al.* in [51] proposed software based *Kernel Fusion* technique that's based on thread interleaved execution. They do not use hardware concurrency and technique achieves suboptimal GPU resource allocation. Wang et al. in [23] have proposed inter-GPU context concurrency technique (before CUDA 4.0 it was not possible). Adriaens et al. in [22] proposed a spatial multitasking method to equally partition streaming multi-processors among the concurrently running kernels based on resource allocation. None of these works explore multi-kernel throughput workload space and lacks systematic power analysis. In [21, 25], the authors have proposed a kernel expansion or contraction framework (*kernel molding*) based on resource requirement. While [21] achieved *kernel molding* only for kernels that can run with an arbitrary number of threads, [25] proposed *elastic kernel* transformation mechanism that supports an arbitrary number of threads for any kernel. However, they did not explore the *kernel affinity* (similarity) score based on the throughput workload space. Moreover, statistical and systematic power–performance co-analysis of throughput workloads based on real GPU data was not considered in either work.

# 7 CONCLUSION

Recent studies have established that a throughput accelerator provides alternative ways to realize exascale computing. Nevertheless, power efficiency requirement to attain exascal computing capability requires extensive architectural enhancement of throughput processors and workload paradigm. By introducing kernel level concurrency in throughput architectures, Nvidia GPUs have further improved the processing throughput and energy efficiency. Naturally, it is pertinent to understand the intrinsic behavior of the enhanced architectural features and its implications for the future of throughput workloads. Unfortunately, such systematic exploration of multi-kernel throughput workloads regarding power and energy efficiency is still lacking. In this paper, we introduce a novel framework for multi-kernel throughput workload generation and perform a thorough study of the proposed workloads concerning performance, power, energy, utilization and interactions between them. Using real Nvidia GPUs of different generations and by varying application scope, we show that power-profile and concurrency are highly correlated. Compared to sequential execution, kernel concurrency improves hardware utilization and helps in reducing the energy footprint of multi-kernel workloads. We present several key findings during this exploration of the concurrent workloads. In summary, the proposed workloads save 32%, 26% and 33% energy on GTX470, Tesla M2050, and Tesla K20 respectively. Using statistical analysis, we demonstrate that proposed workloads possess diversity within the suite.

**Nilanjan Goswami** is senior Graphics Architect at the Facebook Reality Labs. His research interest includes emerging technology based throughput processor design, power-performance co-optimization of throughput core architecture, interconnect, renewable energy based throughput architectures. He has a PhD in the Electrical and Computer Engineering from the University of Florida, Gainesville, FL.

**Yuhai Li** received the B.S. degree, and his Ph.D. degree with the School of Electronic and Information Engineering, Xi'an Jiaotong University in 2008 and 2015. His current research interests include image processing, GPGPU, NoC, and VLSI hardware design for embedded many-core system.

**Amer Qouneh** received the B.S. and M.S. degrees in electrical engineering from Fairleigh Dickinson University, Teaneck, NJ, USA, in 1985 and 1987, respectively; and the M.S. and Ph.D. degrees in computer engineering from the University of Florida, Gainesville, FL, USA, in 2010 and 2014, respectively. He is currently an Assistant Professor with the Department of Electrical and Computer Engineering, Western New England University, Springfield, MA, USA. He was with the Royal Scientific Society, Amman, Jordan, from 1993 to 2000. His current research interests include computer architecture, energy efficiency and power management, power aware scheduling, and high performance computing.

**Chao Li** is a tenure-track assistant professor at Shanghai Jiao Tong University (SJTU). His primary research area is energy-efficient computer architecture and system design. He earned his PhD degree from the University of Florida.

**Tao Li** is a full professor in the Department of Electrical and Computer Engineer- ing at the University of Florida. He received a Ph.D. in Computer Engineering from the Univer- sity of Texas at Austin. His research interest- s include computer architecture, microprocessor/memory/storage system design, virtualiza- tion technologies, energy-efficient/sustainable/ dependable data center, cloud/big data comput- ing platforms, the impacts of emerging technolo- gies/applications on computing, and evaluation of computer systems. Dr. Tao Li received 2009 National Science Foun- dation Faculty Early CAREER Award, 2008, 2007, 2006 IBM Faculty Awards, 2008 Microsoft Research Safe and Scalable Multi-core Computing Award and 2006 Microsoft Research Trustworthy Computing Curriculum Award. Dr. Tao Li co-authored a paper that won the Best Paper Award in HPCA 2011 and three papers that were nominated for the Best Paper Awards in DSN 2011, MICRO 2008 and MASCOTS 2006. Dr. Tao Li is one of the College of Engineering winners, University of Florida Doctor Dissertation Advisor/Mentoring Award for 2013-2014 and 2011-2012.